\documentclass{moriond}

\def\be{\begin{equation}}
\def\ee{\end{equation}}
\def\bea{\begin{eqnarray}}
\def\eea{\end{eqnarray}}

\usepackage{cite}
\usepackage{amssymb}
\usepackage{amsmath}
\usepackage{fdsymbol}

\begin{document}
\vspace*{4cm}
\title{IceCube: Neutrinos from Active Galaxies}

\author{ F. Halzen }
\address{Department of Physics and \\ Wisconsin IceCube Particle Astrophysics Center \\
UW–Madison, Madison, WI 53706 USA}

\maketitle

\abstracts{
The IceCube project transformed a cubic kilometer of transparent, natural Antarctic ice into a Cherenkov detector. It discovered neutrinos of TeV-PeV energy originating beyond our Galaxy with an energy flux that exceeds the one of high-energy gamma rays of extragalactic origin. Unlike at any other wavelength of light, extragalactic neutrinos outshine the nearby sources in our own Milky way. Updated measurements of the diffuse cosmic neutrino flux indicate that the high-energy gamma rays produced by the neutral pions that accompany cosmic neutrinos lose energy in the sources and are likely to be observed at MeV energy, or below. After the reanalysis of 10 years of archival data with an improved data selection and enhanced data analysis methods, the active galaxy NGC 1068 emerged as the hottest spot in the neutrino sky map. It is also the most significant source in a search at the positions of 110 preselected high-energy gamma-ray sources. Additionally, we find evidence for neutrino emission from the active galaxies PKS 1424+240 and TXS 0506+056. TXS 0506+056 had already been identified as a neutrino source in a multimessenger campaign triggered by a neutrino of 290 TeV energy and, by the independent observation of a neutrino burst in 2014 from this source in archival IceCube data. The observations point to active galaxies as the sources of cosmic neutrinos, and cosmic rays, with the gamma-ray-obscured dense cores near the supermassive black holes at their center as the sites where neutrinos originate, typically within $10\sim100$ Schwarzschild radii.}

\section{High-Energy Neutrinos from the Cosmos}
\vspace{.2cm}

The shortest wavelength radiation reaching us from the universe is not radiation at all; it consists of cosmic rays---high-energy nuclei, mostly protons. Some reach us with extreme energies exceeding $10^8$~TeV from a universe beyond our Galaxy that is obscured to gamma rays and from which only neutrinos reach us as astronomical messengers that point back to where they originate~\cite{Aartsen:2013jdh}. The recent observation of neutrinos originating in active galaxies~\cite{IceCube:2018dnn,IceCube:2018cha} and produced in the dense core in the close vicinity of the supermassive black hole~\cite{IceCube:2022der}, represents a breakthrough for resolving the century-old puzzle of the origin of cosmic rays.

\begin{figure}[ht]
\centering
\includegraphics[width=0.6\linewidth, trim=0 150 0 150, clip=true]{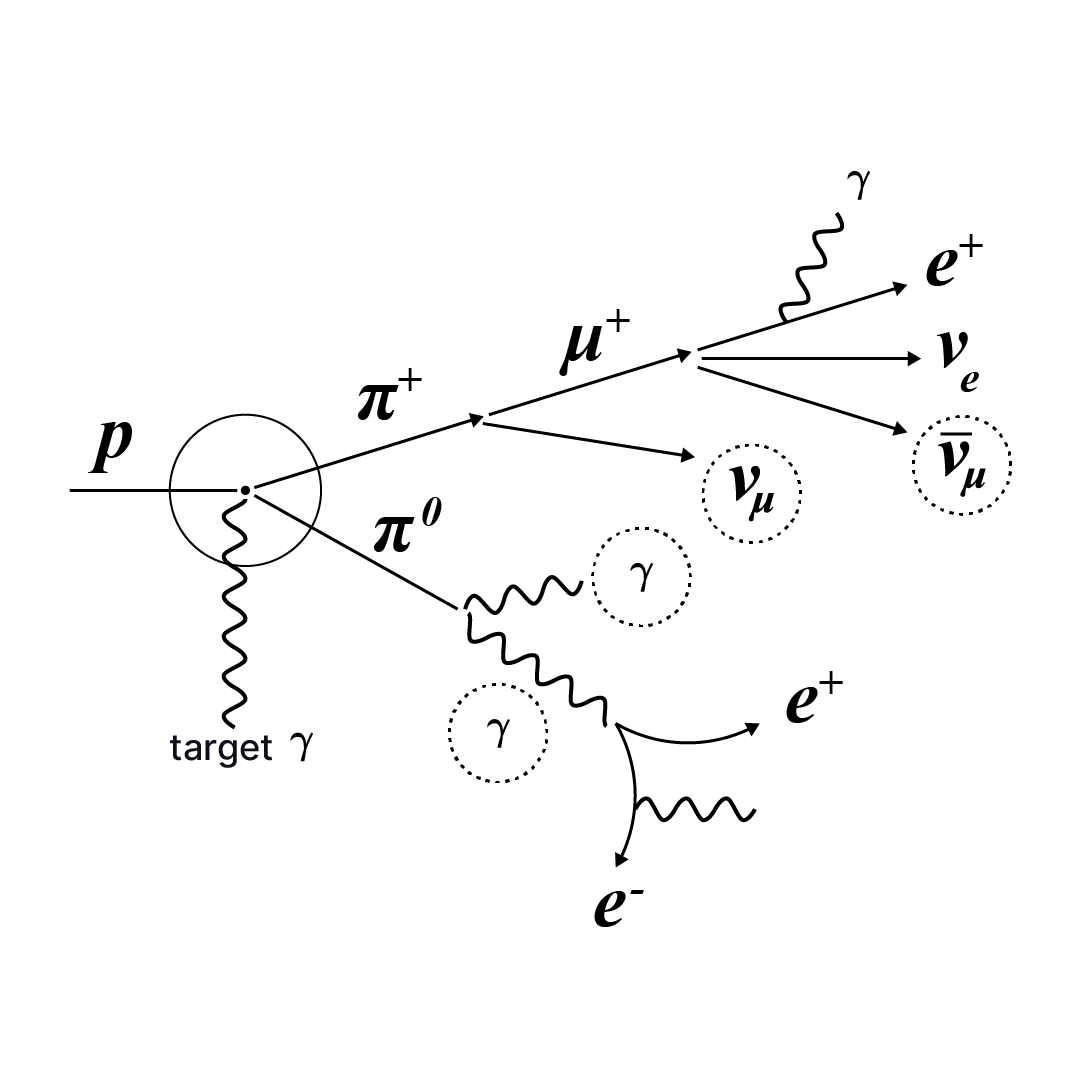}
\caption{Flow diagram showing the production of charged and neutral pions in $p\gamma$ interactions. The circles indicate equal energy going into gamma rays and into pairs of muon neutrinos and antineutrinos, which IceCube cannot distinguish between. Because the charged pion energy is shared roughly equally among four particles and the neutral pion energy among two photons, the photons have twice the energy of the neutrinos.}
\label{fig:flow}
\end{figure}

The rationale for searching for cosmic-ray sources by observing neutrinos is straightforward: in relativistic particle flows, for instance onto black holes, some of the gravitational energy released in the accretion of matter is transformed into the acceleration of protons or heavier nuclei. These subsequently interact with radiation surrounding, or ambient matter accreting onto, the black hole to produce pions and other secondary particles that decay into neutrinos. For instance, when accelerated protons interact with intense radiation fields via the photoproduction processes
\begin{equation}
p + \gamma \rightarrow \pi^0 + p
\mbox{ \ and \ }
p + \gamma \rightarrow \pi^+ + n\,,
\label{eq:delta}
\end{equation}
both neutrinos and gamma rays are produced with roughly equal rates; while neutral pions decay into two gamma rays, $\pi^0\to\gamma+\gamma$, the charged pions decay into three high-energy neutrinos ($\nu$) and antineutrinos ($\bar\nu$) via the decay chain $\pi^+\to\mu^++\nu_\mu$ followed by $\mu^+\to e^++\bar\nu_\mu +\nu_e$; see Fig.~\ref{fig:flow}. Based on this simplified flow diagram, we expect equal fluxes of gamma rays and pairs of $\nu_\mu+\bar\nu_\mu$ neutrinos, which a neutrino telescope cannot distinguish between. Also, from the fact that in the photoproduction process 20\% of the initial proton energy is transferred to the pion\footnote{This is referred to as the inelasticity $\kappa_{p\gamma} \simeq 0.2$}, we anticipate that the gamma ray carries one tenth of the proton energy and the neutrino approximately half of that. Because cosmic neutrinos are inevitably accompanied by high-energy photons, neutrino astronomy is a multimessenger astronomy. While propagating to Earth, unlike neutrinos, the gamma rays interact with microwave photons and other diffuse sources of extragalactic background light (EBL). The $\gamma$-rays lose energy by $e^+e^-$ pair production and the resulting electromagnetic shower subdivides the initial photon energy into multiple photons of reduced energy reaching our telescopes. Importantly, both neutrinos and the accompanying photons trace sources where protons are accelerated, i.e., cosmic-ray sources.

Deployed at a depth of 1450-1550 meter below the geographic South Pole, the IceCube project has transformed a cubic kilometer of transparent Antarctic ice into a Cherenkov detector. After accumulating two years of data, a sample of neutrino events interacting inside the instrumented volume of ice revealed the first evidence for neutrinos of cosmic origin~\cite{Aartsen:2013bka,Aartsen:2013jdh}. Events with PeV energy and no trace of an accompanying muon revealing its atmospheric origin, are highly likely to be of cosmic origin. The present $7.5$-year data sample encompasses a total of 60 neutrino events with deposited energies ranging from 60\,TeV to 10\,PeV. A purely atmospheric explanation of the observation is excluded at $8\sigma$.

\begin{figure*}[ht!]
\centering
\includegraphics[width=0.7\linewidth, trim=0 50 0 80, clip=true]{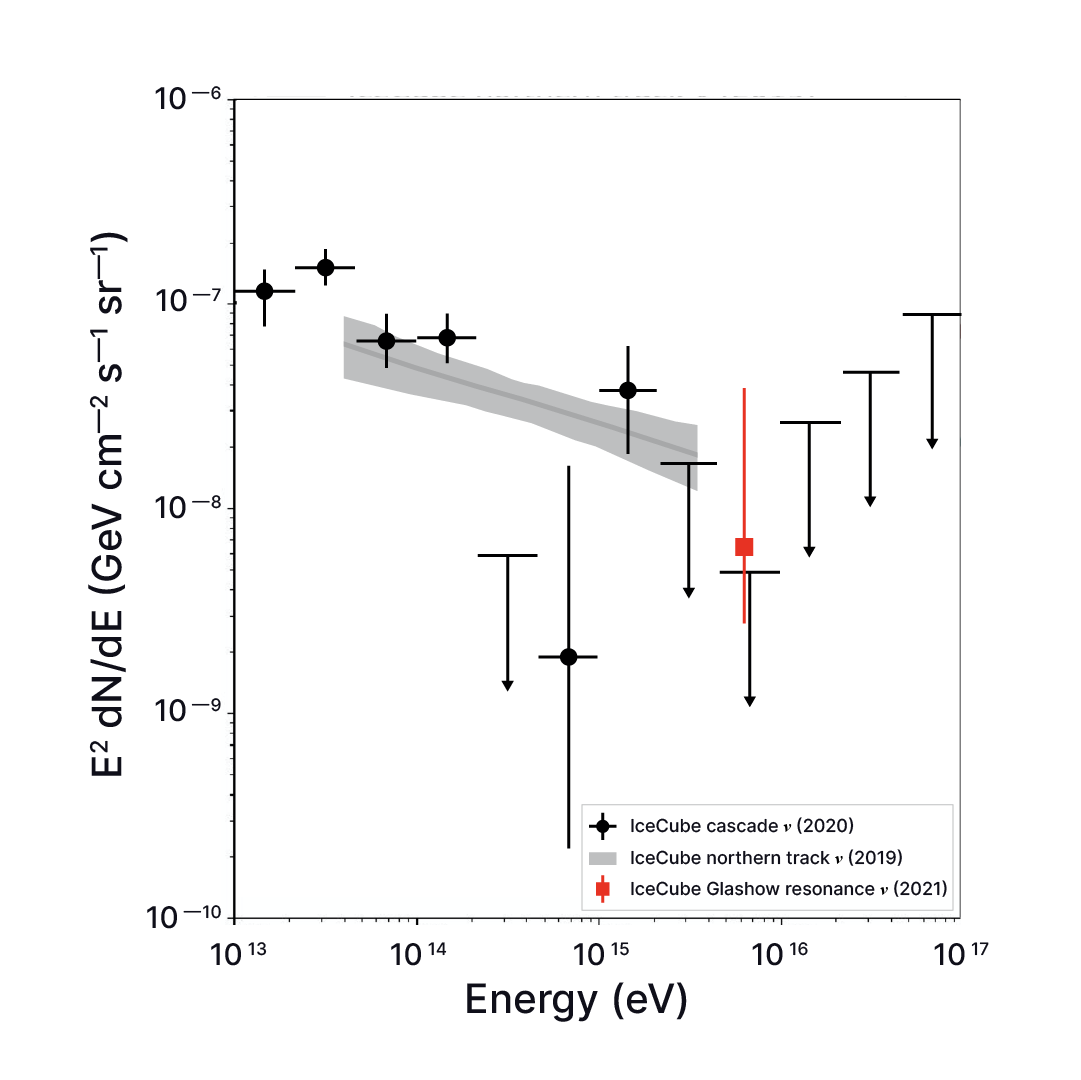}
\caption{The flux of cosmic muon neutrinos~\cite{Aartsen:2017mau} inferred from the $9.5$-year upgoing-muon track analysis (solid line) with $1\sigma$ uncertainty range (shaded) is compared with the flux of showers initiated by electron and tau neutrinos~\cite{IceCube:2020acn}. The measurements are consistent assuming that each neutrino flavor contributes an identical flux to the diffuse spectrum. Both are consistent with the flux derived from the observation of a single Glashow-resonance event.}
\label{fig:showerstracks-2}
\end{figure*}

This discovery has been confirmed using alternative methods to separate the high-energy cosmic neutrinos from the large backgrounds of cosmic ray muons, 3,000 per second, and neutrinos, one every few minutes, produced in the atmosphere. Muon neutrinos of cosmic origin are efficiently identified in samples of muon tracks that are produced by up-going muon neutrinos reaching the South Pole from the Northern Hemisphere, with IceCube using the Earth as a passive shield for the large background of cosmic ray muons. Separating the flux of high-energy cosmic neutrinos from the lower energy neutrinos of atmospheric origin, a spectral index $\rm dN/dE \sim E^{-\gamma}$ is observed with $\gamma = -2.37\pm0.09$ above an energy of $\sim 100$\,TeV~\cite{Aartsen:2017mau}; see Figure~\ref{fig:showerstracks-2}. This value is somewhat larger than the value of $-2.87\pm0.2$ for the 68\% confidence interval, obtained for events starting inside the detector~\cite{Abbasi_2021}. Figure~\ref{fig:showerstracks-2} also shows the results of yet another search that exclusively identifies showers initiated by electron and tau neutrinos that can be isolated from the atmospheric background to energies as low as 10\,TeV~\cite{IceCube:2020fpi}. Reaching us from long baselines the cosmic neutrinos flux has oscillated to a ratio $\nu_e:\nu_\tau:\nu_\mu$ of approximately 1:1:1.
\vspace{-5pt}

Besides the three independent analyses discussed, IceCube has confirmed the existence of neutrinos of cosmic origin by the observation of high-energy tau neutrinos~\cite{Abbasi:2020zmr} and, by the identification of a Glashow resonant event. Identified in a dedicated search for partially contained showers of very high energy, the reconstructed energy of the Glashow event is 6.3\,PeV, matching the laboratory energy for the production of a weak intermediate $W^-$ boson in the resonant interaction of an electron antineutrino with an atomic electron~\cite{IceCube:2021rpz}: $\bar{\nu}_e + e^- \rightarrow W^- \rightarrow q + \bar{q}$. Given its high energy, the initial neutrino is cosmic in origin and provides an independent discovery of cosmic neutrinos at the level of $5.2\sigma$.

We can now calculate the flux of gamma rays produced by the neutral pions that accompany the neutrino flux in Fig.~\ref{fig:showerstracks-2}. After propagating the photons through the EBL to our telescopes at Earth, we find that it exceeds the diffuse extragalactic photon flux observed, for instance by the NASA Fermi satellite. There is no contradiction here; we infer from the calculation that the pionic gamma rays already lose energy in the target producing the neutrinos, and not just in the EBL, and emerge below the threshold of the instrument, at MeV energies or below. We conclude that the observed diffuse neutrino flux points at $\gamma$-ray-obscured sources; see Ref.~\citeonline{Fang:2022trf} for a recent analysis. This conclusion will be confirmed by the observation of the first sources.

\section{Observation of the First Neutrino Sources}
\vspace{.2cm}

Starting with different configurations of the partially completed detector, IceCube has performed searches for sources of high-energy neutrinos by studying the arrival directions of well-reconstructed muon neutrinos from the Northern Hemisphere using the Earth as a filter. Using long-established maximum likelihood methods~\cite{BRAUN2008299}, we search the sky for clusters of high-energy neutrinos. Additionally, having in mind the multimessenger interface between pionic photons and neutrinos, we search in the directions of preselected known gamma-ray emitters, 110 in the most recent analyses.

The most recent search of the neutrino sky has revealed three sources~\cite{IceCube:2022der}: the active galaxies NGC 1068, PKS 1424+240, and TXS 0506+056. The first, NGC 1068 (M77), is the most significant source in both searches mentioned above. The active galaxy TXS 0506+056 had already been identified as a neutrino source in a multimessenger campaign triggered by a neutrino of 290 TeV energy, IC170922, as well as by the independent observation of a burst of neutrinos from the same source in archival data in 2014~\cite{IceCube:2018dnn,IceCube:2018cha}.

Identified as powerful particle accelerators by astronomers, the emergence of active galaxies as the first sources of cosmic rays is not unexpected. However, a neutrino source is more than an accelerator, it requires a $\it proton$ beam as well as a target converting protons into pions that decay into neutrinos. In some active galaxies, the radiation and accreting matter near the black hole reach extremely high densities in a very small volume referred to as the corona, providing a dense target for producing neutrinos~\cite{Murase:2019vdl}. The emergence of these, and other sources observed with diminishing significance~\cite{IceCube:2022der}, suggest that the neutrinos may originate in the gamma-ray-obscured cores of active galaxies, typically within $10\sim100$ Schwarzschild radii of the supermassive black hole. We will return to this fascinating observation at the conclusion of this review.

\begin{figure}[ht]
\centering
\includegraphics[width=\columnwidth]{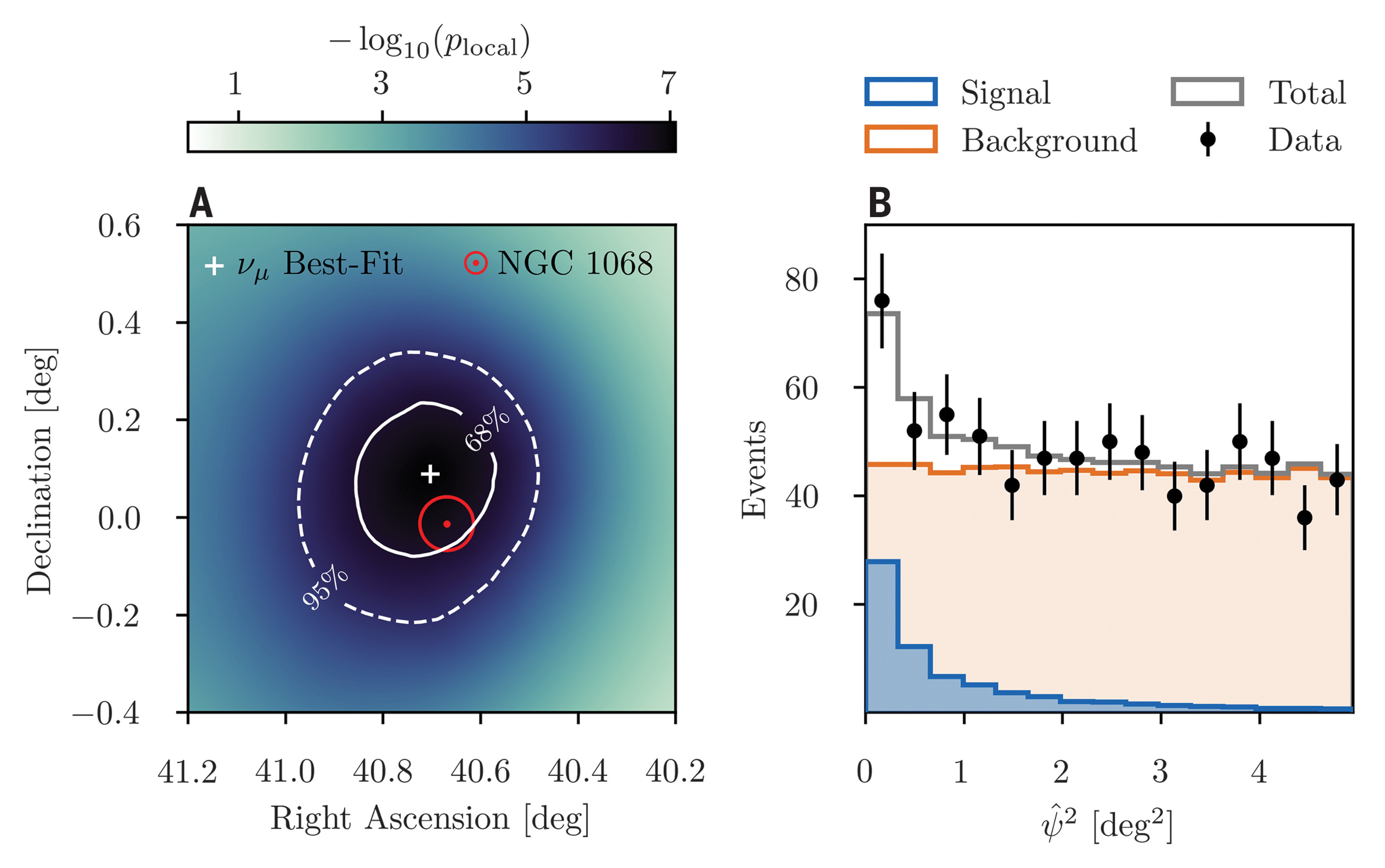}
\caption{High-resolution scan around the most significant location in the sky map. (a) High-resolution scan around the most significant location marked by a white cross, with contours showing its 68\% (solid) and 95\% (dashed) confidence regions. The red dot shows the position of NGC 1068, and the red circle is its angular size in the optical wavelength. (b) The distribution of the angular distance squared, $\Psi^ 2$, between NGC 1068 and the reconstructed event directions. We estimated the background (orange) and the signal (blue) from Monte Carlo simulations, assuming the best-fitting spectrum at the position of NGC 1068. The superposition of both components is shown in gray and the data in black.}
\label{fig:ngczoom}
\end{figure}

With the observation of NGC 1068 approaching the level of evidence at $2.9\sigma$~\cite{Aartsen:2019fau} after 10 years, IceCube initiated a campaign to update the detector calibrations in combination with improved modeling of the optics of the ice and the development of superior methods for reconstructing the direction and energy of neutrinos using the more modern machine learning tools that have become recently available. Using archival data we reconstructed a sample of 670,000 muon neutrinos with a purity of 99.7\% covering the declinations from $-3^{\circ}$ to $81^{\circ}$. Searching the entire sky resulted in a hot spot with a local signficance of $5.3\sigma$, which is however reduced to $2.0\sigma$ after accounting for all trials in the search. A refined scan zooming in on its location reveals NGC 1068 within $\sim 0.18^{\circ}$, i.e., within the IceCube resolution; see Fig.~\ref{fig:ngczoom}.

\begin{figure}[ht]
\centering
\includegraphics[width=\columnwidth]{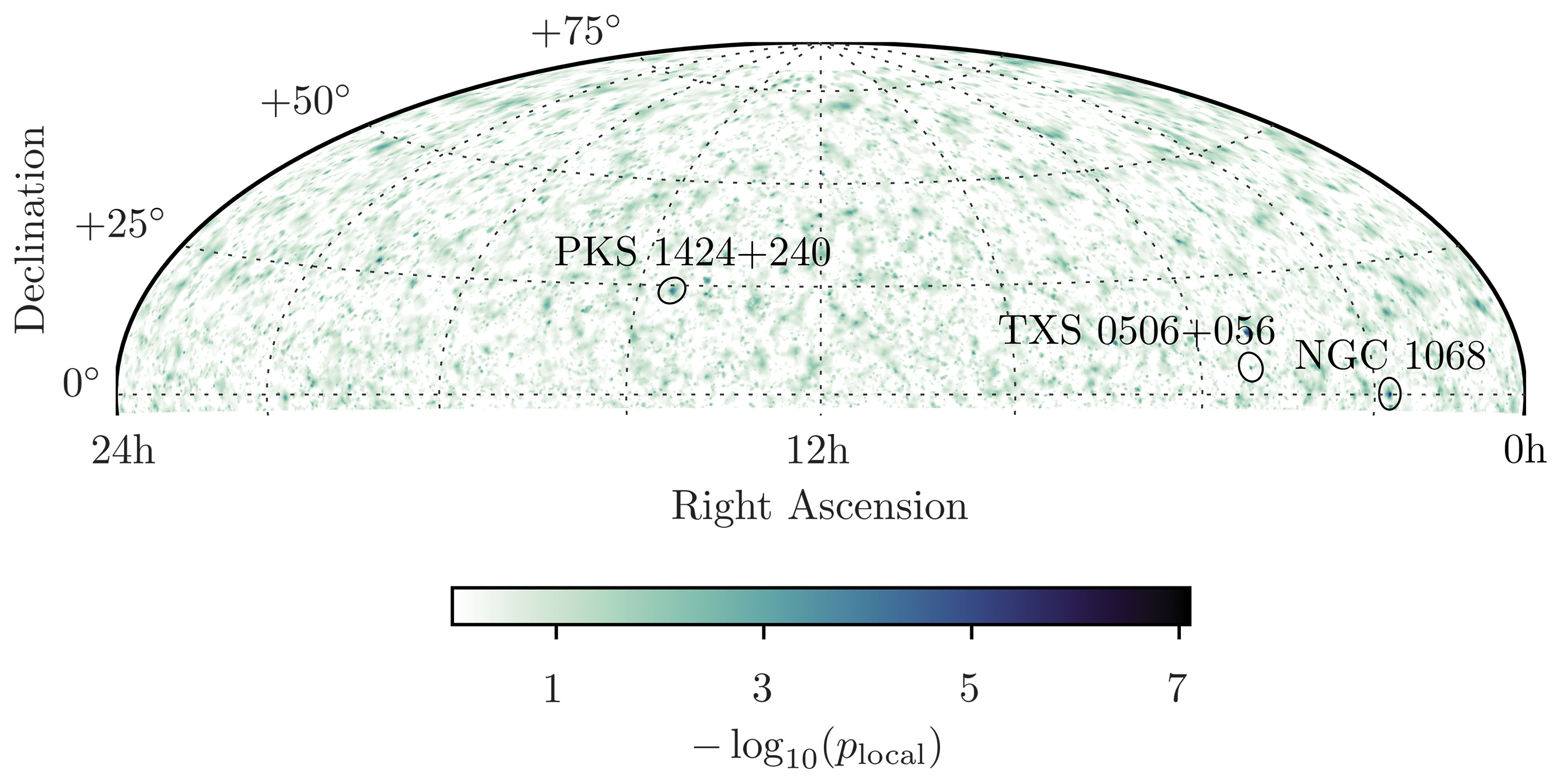}
\caption{Sky map of the scan for point sources in the Northern Hemisphere. The color scale indicates the logarithm 
of the local P value (Plocal) obtained from the maximum likelihood analysis, evaluated at each location in the sky. The spectral index is left free as a parameter. The map is shown in equatorial coordinates. The black circles indicate the three most significant objects in the source list search. The circle around NGC 1068 contains the most significant location in the Northern Hemisphere, shown in higher resolution in Fig.~\ref{fig:ngczoom}.}
\label{fig:skymap}
\end{figure}

We subsequently search at the locations in the sky of preselected astronomical objects in the Northern Hemisphere. It is important to point out here that we performed a blind analysis with all methods, including the selection of the hypotheses to be tested, formulated a priori, i.e., all trials can be accounted for. Of the 110 astronomical objects tested, NGC 1068 is also the most significant here, with a local P value of 1 × $10^{-7}$, or $5.2\sigma$; its flux has best-fitting values of spectral index of $3.2 \pm 0.2$ and mean number of signal events of $79^{+22}_{-20}$. NGC 1068 is contained within the 68\% confidence region around the most significant location in the sky scan, offset by 0.11°, consistent with neutrino emission from NGC 1068 (Fig.~\ref{fig:ngczoom}). After correcting for having tested the 110 sources in the catalog, the global P value for NGC 1068 is 1.1 × $10^{-5}$, corresponding to a significance of $4.2\sigma$.

Both scans were performed assuming a power-law neutrino flux with the spectral index as a free parameter. They were subsequently repeated for spectral indices fixed at 2.0 and 2.5. It is intriguing that NGC 4151 is the fourth source in the latter analysis; it was, with NGC 1068, one of the two brightest spirals detected by Carl Seyfert in 1943~\cite{Seyfert:1943}.

A binomial scan of the list of sources reveals that three dominant sources contribute an excess with a combined significance of $3.4\sigma$ after correcting for all trials: NGC 1068, and two additional sources PKS 1424+240 and TXS 0506+056, with local significances of $3.7\sigma$ and $3.5\sigma$, respectively. These were already the dominant sources in a previous analysis ~\cite{Aartsen:2019fau}.

The $3.5\sigma$ evidence for TXS 0506+056 in a time-integrated analysis confirms the previous identification of this source by its transient emission in 2014~\cite{IceCube:2018cha} and in a multimessenger campaign in 2017~\cite{IceCube:2018dnn}. The case that its observation is compelling has been made in detail in Ref.~\citeonline{Halzen:2021xkf}; we briefly summarize it here.

Within 43 seconds of a neutrino of energy 290\,TeV, IC170922, stopping in the instrumented Antarctic ice, its arrival direction was reconstructed and automatically sent to the Gamma-ray Coordinate Network (GCN) for potential follow-up by astronomical telescopes. Its arrival direction was aligned with the coordinates of a known Fermi ``blazar," TXS 0506+056, to within $0.06^\circ$. The source was ``flaring," with a gamma-ray flux that had increased by a factor of seven in recent months. A variety of estimates converged on a probability on the order of $10^{-3}$ that this coincidence was accidental. The identification of the neutrino with the source reached the level of evidence, but not more. What clinched the association were observations following the multimessenger campaign.

Informed where to look, IceCube searched its archival neutrino data up to and including October 2017 for evidence of neutrino emission at the location of TXS 0506+056~\cite{IceCube:2018cha}. Evidence was found in 2014-15 for 19 high-energy neutrino events on a background of fewer than 6 in a burst lasting 110 days. This burst dominates the integrated flux from the source over the last 9.5 years of archival IceCube data, leaving the 2017 flare as a second subdominant feature. This observation is consistent with the evidence for the source found in the time-independent analysis discussed above. Interestingly, the burst is not accompanied by high-energy gamma rays.

The association of IC170922 with TXS 0506+056 has been sealed by an optical observation, initiated 73 seconds after the alert, that observed the source switching from an ``off" to ``on" state two hours after the emission of IC-170922A, conclusively associating the neutrino with TXS 0506+056 in the time domain~\cite{2020ApJ...896L..19L}. The rapid transition resulted in a doubling of its optical luminosity~\cite{2020ApJ...896L..19L}. Such optical flares of active galaxies are often associated with magnetohydrodynamical instabilities triggered by processes modulated by the magnetic field of the accretion disk. This may hint at the production of neutrinos near the core, as is the case for NGC 1068.
 
\section{Do Cosmic Neutrinos Originate in the Obscured Cores of Active Galaxies?}
\vspace{.2cm}

\begin{figure}[ht!]
\centering
\includegraphics[width=0.7\columnwidth]{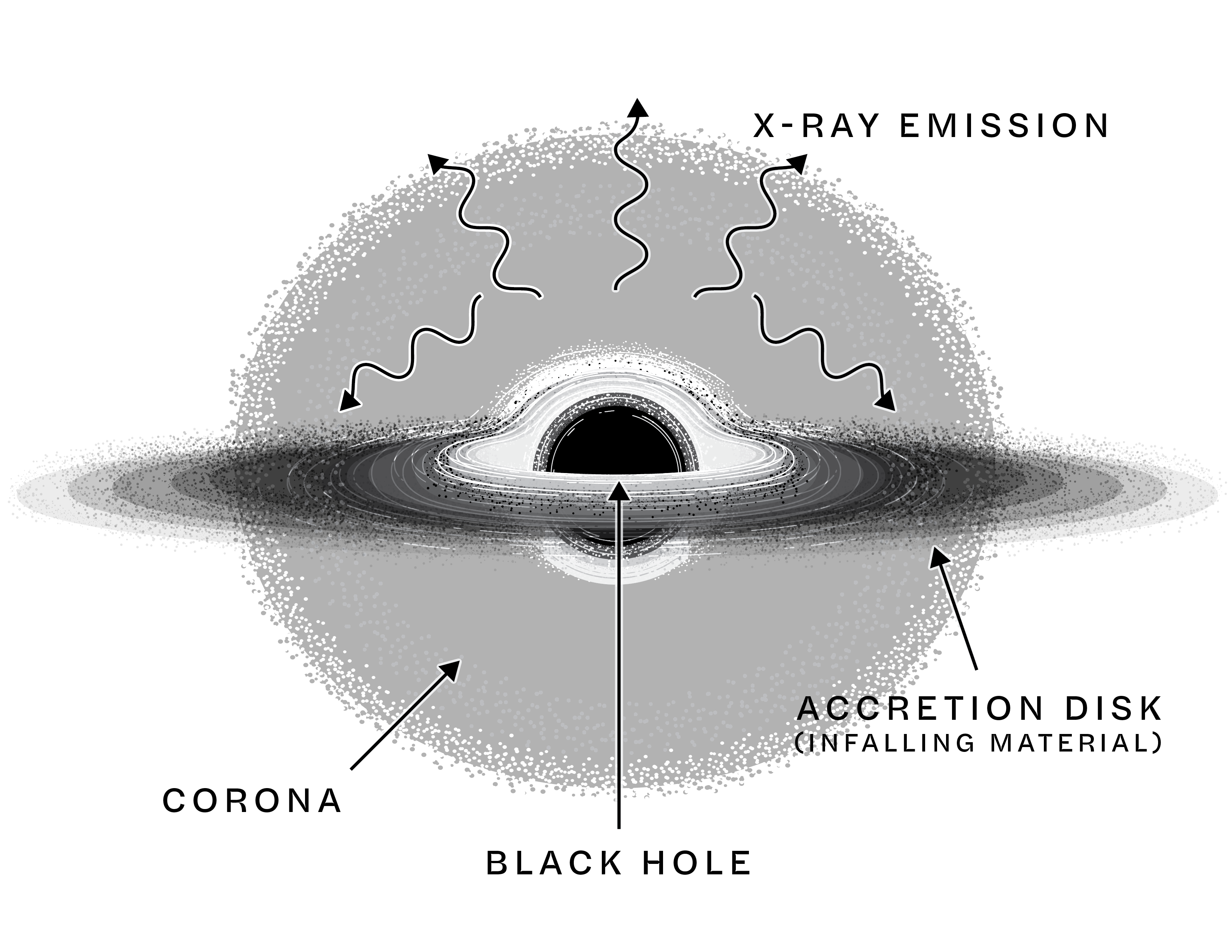}
\caption{The core of an active galaxy combines the ingredients to produce neutrinos, with protons accelerated by the black hole and/or the accretion disk and a target (corona) with a large optical depth in both gamma rays and hydrogen that converts them into neutrinos.}
\label{fig:Black-Hole_Diagram}
\end{figure}

With the accumulating evidence that neutrinos originate in $\gamma$-ray obscured sources, I would like to emphasize that this should not come as a surprise. Imagine that protons, accelerated near the black hole or in the accretion disk, interact in a photon target of size $R$ centered on the black hole; see Fig.~\ref{fig:Black-Hole_Diagram}. Astronomers refer to such structures with large densities in both X-rays and accreting matter as a ``corona." The opacity of such a target to accelerated protons\footnote{The opacity is actually the energy-loss length and $\tau$ should be replaced by $1-e^{-\tau}$ whenever it is large~\cite{Halzen:2022pez}.} is
\begin{equation}
\label{eq:pgammaopacity}
\rm \tau_{p\gamma}\simeq \frac{\kappa_{p\gamma} R_{escape}}{\lambda_{p\gamma}} \simeq \kappa_{p\gamma} \rm R_{escape}\, \sigma_{p\gamma} \,n_{\gamma},
\end{equation}
which is determined by how often the proton interacts in a target of size $\rm R_{target}$ given its interaction length $\lambda_{p\gamma}$. With each interaction, the proton loses a fraction $\kappa_{p\gamma}$ of its energy, the inelasticity. The interaction length is determined by the density of target photons $n_\gamma$ and the interaction cross section $\sigma_{p\gamma}$\footnote{For the simple dimensional analysis in this section we use the following cross sections $\rm \sigma_{\gamma\gamma} = 6.7 \times 10^{-25} cm^2$, $\rm \sigma_{p\gamma} = 5 \times 10^{-28} cm^2$, and $\rm \sigma_{pp} = 3 \times 10^{-26} cm^2$.}.

The opacity of the target to the photons produced along with the neutrinos is given by
\begin{equation}
\tau_{\gamma\gamma} \simeq \rm R_{target}\, \sigma_{\gamma\gamma}\,n_{\gamma},
\end{equation}
and therefore, approximately, the two opacities are related by their cross sections 
\begin{equation}
\tau_{\gamma\gamma} \simeq \frac{\sigma_{\gamma\gamma}}{\kappa_{p\gamma} \sigma_{p\gamma}} \, \tau_{p\gamma} \simeq 10^3\,  \tau_{p\gamma}
\label{opacitygamma}
\end{equation}
for $\rm R_{escape} \sim R_{target} \sim R$. A target that produces neutrinos with $\tau_{p\gamma} \gtrsim 0.1$ will not be transparent to the pionic gamma rays, which will lose energy in the target even before propagating in the EBL. For instance, we should not expect neutrinos to be significantly produced in blazar jets that are transparent to very high energy gamma rays. In contrast, the highly obscured dense cores close to supermassive black holes in active galaxies represent an excellent opportunity to produce neutrinos, besides providing opportunities for accelerating protons.

Pursuing these dimensional arguments for the X-ray emitting corona of an active galaxy, we introduce a conceptual version of the neutrino source in Fig.~\ref{fig:Black-Hole_Diagram} showing the core of the active galaxy obscured by a dense corona of radius $\rm R$ that converts accelerated protons into neutrinos. We estimate its opacity to protons from Eq.~\ref{eq:pgammaopacity} with a target density $\rm n_{\gamma} \simeq n_X$, with
\begin{equation}
\rm n_X = \frac{u_X}{E_X} = \frac{1}{4 \pi R^2c}\, \frac{L_X}{E_X}\,,
\end{equation}
and therefore
\begin{equation}
\rm \tau_{p\gamma} = \frac{\kappa_{p\gamma} \sigma_{p\gamma}}{4 \pi c}\, \frac{1}{R}\, \frac {L_X}{E_X}
\end{equation}
We here determined the target density in X-rays from the energy density of X-rays $\rm u_X$ divided by their energy $\rm E_X$ and, subsequently, identified the energy flux $\rm cu_X$ with the measured luminosity of the source $\rm 4 \pi R^2\, L_X$. In the end the opacity of the corona to protons is proportional to the density of target photons, i.e., to the X-ray luminosity $\rm L_X$, and inversely proportional to the radius $\rm R$. 

In order to study the implications, we rewrite the opacity in terms of the Schwarzschild radius and the Eddington flux of the supermassive black, with a mass of about $\rm 10^7\, M_{sun}$ for NGC 1068:
\begin{equation}
\rm R_s = [2G\rm M]/c^2 = 3\times10^5 \rm cm \,[\frac{\rm M}{\rm M_{sun}}]\,,
\end{equation}
and the Eddington luminosity by
\begin{equation}
\rm L_{\rm edd} = 4 \pi G {\rm M} m_p c / \sigma_{\rm T} = 1.2 \times 10^{38} \frac{\rm erg}{\rm s}\,[\frac{\rm M}{\rm M_{sun}}].
\end{equation}
$\rm M$ is the mass of the supermassive black hole, G Newton's constant, $c$ the speed of light, $m_p$ the proton mass, and $\sigma_T$ the Thomson cross section. We thus obtain the result that
\begin{equation}
\label{eq:targetopacity}
\rm \tau_{p\gamma} \simeq 10^2\,[\frac{R_s}{R}]\,[\frac{1\,\rm keV}{E_X}][\frac{L_X}{L_{\rm edd}}] \simeq 10^{-3}\,\tau_{\gamma\gamma}.
\end{equation}
\begin{figure}[ht!]
\centering
\includegraphics[width=0.7\columnwidth]{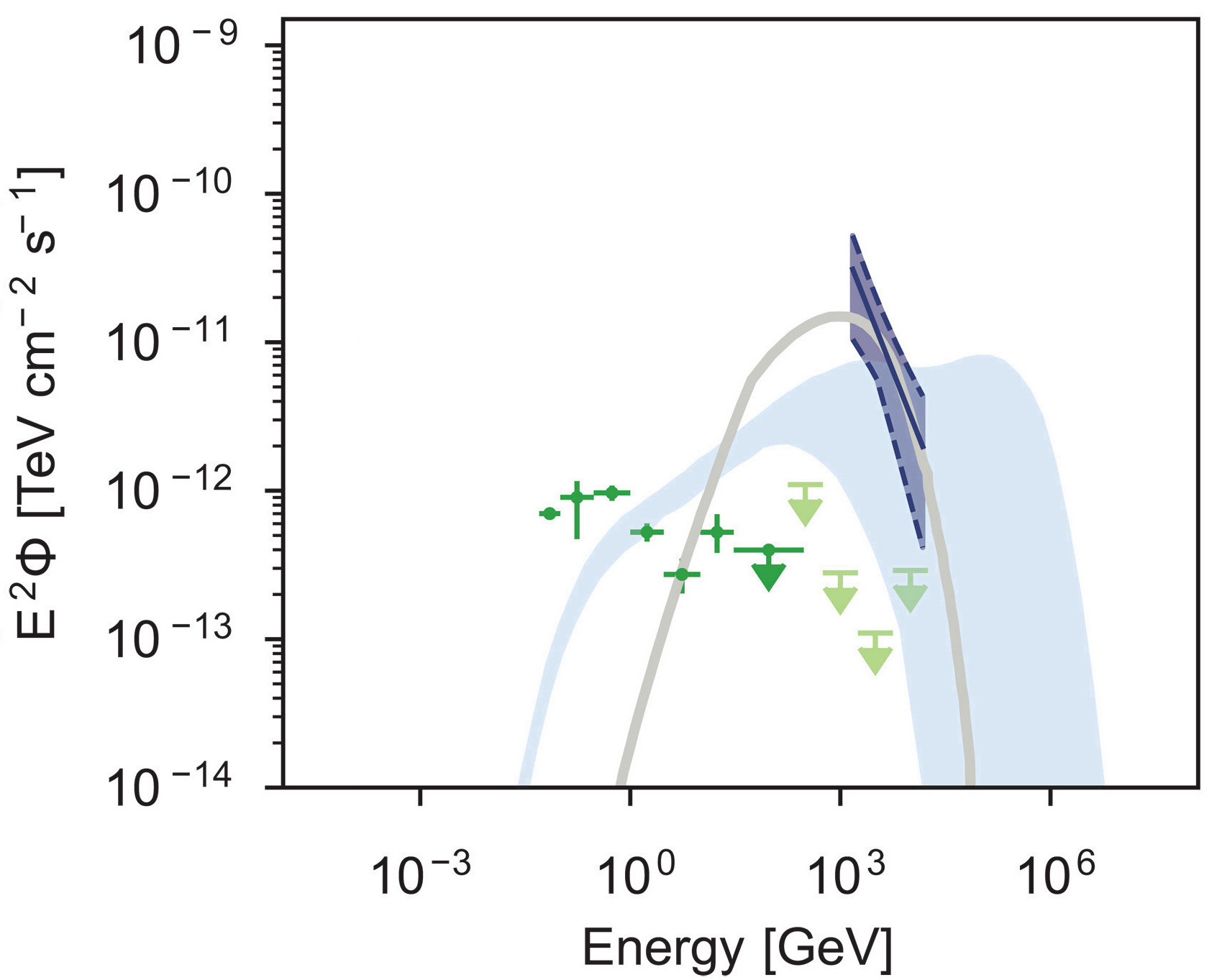}
\caption{Multimessenger spectral energy distribution of NGC 1068. Dark and light green points indicate gamma-ray observations at 0.1 to 100 GeV~\cite{Fermi-LAT:2019pir,Abdollahi_2020} and $>200$ GeV~\cite{Acciari:2019raw}, respectively. Arrows indicate upper limits, and error bars are $1\sigma$ confidence intervals. The solid, dark blue line shows our best-fitting neutrino spectrum, with the dark blue shaded region indicating the 95\% confidence region. We restrict this spectrum to the range between 1.5 and 15 TeV, where the flux measurement is well constrained (26). Two theoretical predictions are shown for comparison: The light blue shaded region and the gray line show the NGC~1068 neutrino emission models from Refs.~\citeonline{Inoue2019abc}, \citeonline{Inoue:2019yfs}, and~\citeonline{Murase:2019vdl}, respectively. All fluxes are multiplied by the energy squared $\rm E^2$.}
\label{fig:ngcdatatheory}
\end{figure}

The neutrino flux from NGC 1068 is shown in Fig.~\ref{fig:ngcdatatheory} along with the multiwavelength gamma ray data. Note the MAGIC upper limits that exclude the presence of the accompanying TeV flux by more than one order of magnitude; the source is gamma-ray obscured. In order to produce the neutrinos we require a relatively large opacity to protons in the target, $\tau_{p\gamma} \gtrsim 0.1$. Importantly, this will automatically guarantee the suppression of the pionic photon flux by more than one order of magnitude relative to neutrinos by Eq.~\ref{opacitygamma}. For NGC 1068, $\rm L_X \sim 10^{43} \,\rm erg/s$ at keV energy, or $\rm L_X \simeq 10^{-2}\, L_{\rm edd}$. This implies that $\rm R \simeq 10 \, R_s$. Detailed modeling of the accelerator and target reinforce our conclusion that neutrinos originate within $\rm \sim 10^{-4}\,pc$ of the supermassive black hole. Producing neutrinos, even at the base of the jet at a distance within parsecs of the black hole, will fail to yield the neutrino flux observed for NGC 1068 and suppress the appearance of a TeV gamma ray flux at TeV energy.

Interestingly, high densities of both radiation~\cite{Bauer_2015,Marinucci:2015fqo,Ricci_2017} and matter~\cite{Rosas:2021zbx,Garcia-Burillo_2016} are associated with the core of NGC 1068 and therefore neutrinos are efficiently produced in both $p\gamma$ and $pp$ interactions. A second look of the multiwavelength data shown in Fig.~\ref{fig:ngcdatatheory} leads us to conclude that $\tau_{pp} > \tau_{p\gamma}$ because, while protons are absorbed by target photons through both $p\gamma$ photoproduction and the $p\gamma \rightarrow pe^+e^-$ Bethe-Heitler process, the $p\gamma$ threshold for producing neutrinos on a target of keV-energy photons at the $\Delta$-resonance, is at PeV energy. Instead, the typical energy of the neutrinos observed by IceCube from the direction or NGC 1068 is closer to $\sim 10$\,TeV; see Fig.~\ref{fig:ngcdatatheory}. These neutrinos are predominantly produced in $pp$ interactions with a reduced threshold for producing multiple pions that result in neutrinos of lower energy.

From Eq.~\ref{eq:pgammaopacity} we calculate the opacity of the target on the large matter density close to the black hole
\begin{equation}
\rm \tau_{pp} \simeq \frac{\kappa_{pp}\, R} {\lambda_{pp}} \simeq \kappa_{pp}\, \rm R\, \sigma_{pp} \,n_p \simeq \kappa_{pp}\, \sigma_{pp}\, N_H,
\label{eq:gas}
\end{equation}
where we have introduced the line-of-sight density $\rm N_H = R\, n_p$, which is sufficiently large for NGC 1068 to achieve an opacity $\rm \tau_{pp} \gtrsim 1$. In Fig.~\ref{fig:ngcdatatheory}, two such models are compared to the flux observed by IceCube from NGC 1068. Accelerated protons interact in the corona with both the dense gas in $pp$ interaction and with the X-ray photons by $p\gamma \rightarrow pe^+e^-$, the Bethe-Heitler process.

From Eq.~\ref{eq:gas}, one may be tempted to conclude that, with the production of neutrinos in $pp$ interactions only depending on the line-of-sight density $N_H$, we can evade the constraint on the size of the production region $\rm R \lesssim 100 \, R_s$. This is not the case because a high density of X-rays in the corona is still required to suppress the flux of pionic gamma rays produced in $pp$ interactions in order to evade the strong upper limits from the the MAGIC telescope shown in Fig.~\ref{fig:ngcdatatheory}. The suppression of the $\gamma$-ray flux by over one order of magnitude cannot be achieved with interactions with protons only. Although produced by different mechanisms, the gamma rays accompanying the neutrinos still emerge at MeV energies, the characteristic energy at which the X-ray corona becomes transparent to photons.

Detailed modeling of the multiwavelength spectrum is more challenging with the GeV-energy gamma ray flux shown in Fig.~\ref{fig:ngcdatatheory} pointing at other low-energy contributions, possibly from the star-forming region in NGC 1068 or the inverse Compton scattering of nonthermal electrons. Recently, an analysis of the neutrino data by Ke Fang and collaborators of the circumnuclear region of NGC 1068 suggests that the region where the torus and jet interact with the interstellar medium may also be a potential site for producing neutrinos (in preparation). For a more detailed review of the models, see Refs.~\citeonline{Murase:2022dog,Kheirandish:2021wkm,Eichmann:2022lxh,Anchordoqui:2021vms}.

It is intriguing~\cite{Anchordoqui:2021vms} that the number of active galaxies with X-ray flux exceeding $10^{43} \rm erg/s$ is $\rm n \sim 10^3 \,\rm Gpc^{-3}$~\cite{Urry:1995mg}; these can collectively reproduce the diffuse neutrino flux observed by IceCube:
\begin{equation}
\rm \Phi_\nu \simeq \frac{1}{4\pi}\, R_H\, n\, L_\nu\,,
\end{equation}
with $\rm L_\nu \sim 10^{-2}L_XE_\nu^2$ and $\rm R_H$ the Hubble radius. Reinforced by the fact that the energies of cosmic rays and neutrinos in the universe are similar, this represents a possible blueprint for the solution of the problem of the origin of cosmic rays.

The emergence of active galaxies as cosmic ray sources coincides with the demise of gamma-ray bursts (GRBs)~\cite{IceCube:2012qza}. The strong limits obtained by IceCube on the neutrino emission from the most powerful GRB ever observed~\cite{Abbasi:2023xhh} have recently further constrained their contribution to the cosmic ray spectrum~\cite{Murase:2022vqf}. 

Do cosmic neutrinos (and cosmic rays) originate in the obscured cores of active galaxies? Some do, and with NGC 1068, we have identified the first source of high-energy cosmic rays since their discovery more than a century ago. Instead, the modeling of the multiwavelength spectrum of TXS 0506+056 represents an unmet challenge. The source is different: it emits neutrinos in bursts with Eddington luminosity characterized by a harder neutrino spectrum, close to $\rm E^{-2.2}$. The gamma-ray obscured spectra at the time of neutrino production~\cite{Halzen:2021xkf} may indicate that the neutrinos are produced in the vicinity of the core in association with catastrophic events hinted at by the observation of a strong optical flare in 2017~\cite{2020ApJ...896L..19L}.

An inescapable inference of this review is that progress will require more neutrinos with better angular resolution. As a next step, IceCube has produced a technical design for a next-generation detector instrumenting more than $8\rm\,km^3$ of glacial ice at the South Pole, capitalizing on the large absorption length of Cherenkov light in ice~\cite{IceCube-Gen2:2020qha}. In the meantime, neutrino telescopes are under construction in Lake Baikal ~\cite{Shoibonov:2019gfj}, in the Mediterranean Sea~\cite{Sanguineti:2023qfa} and, in the Pacific Ocean~\cite{Bailly:2021dxn}. In China, two initiatives are in the planning stages~\cite{Ye:2022vbk}, one contemplating the construction of a $30\,\rm km^3$ telescope.

\section{Acknowledgements}

Discussion with collaborators inside and outside the IceCube Collaboration, too many to be listed, have greatly shaped this presentation. Thanks as well to Haotian Cao, Khatee Zathul Arifa, Ali Kheirandish, Qinrui Liu and Marjon Moulai for a reading of the manuscript.

This research was supported in part by the U.S. National Science Foundation under grants~PLR-1600823 and PHY-1913607 and by the University of Wisconsin Research Committee with funds granted by the Wisconsin Alumni Research Foundation.

\section*{References}

\bibliography{bib}





\end{document}